\documentclass[12pt]{article}

\ifx\pdfoutput\undefined
\usepackage[dvips,bookmarks]{hyperref}
\else
\usepackage{hyperref}
\fi
\hypersetup{colorlinks=false,bookmarksopen,bookmarksnumbered,citecolor=blue,
   pdfstartview=FitH}

\usepackage{latexsym}

\usepackage{amssymb,amsfonts,amsmath}
\usepackage{graphicx}
\usepackage{indentfirst}

 \usepackage{bbm}

\parskip=\medskipamount

\def\a{\alpha}

\def\b{\beta}
\def\c{\chi}
\def\d{\delta}
\def\e{\epsilon}
\def\f{\phi}
\def\g{\gamma}
\def\G{\Gamma}

\def\k{\kappa}
\def\l{\lambda}
\def\m{\mu}
\def\n{\nu}

\def\p{\pi}
\def\q{\theta}

\def\s{\sigma}

\newcommand{\vf}{\varphi}

\newcommand{\be}{\begin{equation}}
\newcommand{\ee}{\end{equation}}
\newcommand{\bea}{\begin{eqnarray}}
\newcommand{\eea}{\end{eqnarray}}

\def\double #1{#1{\hbox{\kern-2pt $#1$}}}

\newcommand{\W}{\Omega}
\newcommand{\no}{\nonumber}
\newcommand{\ba}{\bar}
\newcommand{\ii}{\mathrm{i}}
\newcommand{\ex}{\mathrm{e}}

\newcommand{\bsubeq}{\begin{subequations}}
\newcommand{\esubeq}{\end{subequations}}

\begin{document}

\begin{titlepage}

\vspace{2cm}

\begin{center}
{\Large \bf  $p$-brane superalgebras via integrability }
\end{center}

\vspace{2cm}

\begin{center}

{\large

D. T. Grasso\footnote{{darren.grasso@uwa.edu.au}},
I. N. McArthur\footnote{{ian.mcarthur@uwa.edu.au}}
} \\
\vspace{5mm}

\footnotesize{
{\it School of Physics M013,
The University of Western Australia\\
35 Stirling Highway, Crawley W.A. 6009 Australia }}
~\\

\end{center}
\vspace{3cm}

\begin{abstract}
\baselineskip=14pt
\noindent
It has long been appreciated that superalgebras with bosonic and fermionic generators additional to those in the super-Poincare algebra underlie $p$-brane and $D$-brane actions in superstring theory. These algebras have been revealed via ``bottom up'' approaches, involving consideration of Noether charges, and by ``top down'' approaches, involving the construction of manifestly supersymmetry invariant Wess-Zumino actions. In this paper, we give an alternative derivation of these algebras based on integrability of supersymmetry transformations assigned to fields in order to solve a cohomology problem related to the construction of Wess-Zumino terms for $p$-brane and $D$-brane actions.
\end{abstract}
\vspace{1cm}

\vfill
\end{titlepage}

\tableofcontents{}
\vspace{1cm}
\bigskip\hrule

\section{Introduction}
\setcounter{equation}{0}
One of the major breakthroughs in superstring theory came with the realisation that in addition to strings, world-volume structures of higher  dimension  ($p$-branes and $D$-branes) contribute to a  much richer spectrum, and indeed are integral to dualities that relate the seemingly different superstring theories.

For $p$-branes or  $D$-branes embedded in flat superspace, it has long been appreciated that enlarged\footnote{We avoid the term ``extended," as extended supersymmetry algebras involve multiple supersymmetry generators; in this paper we consider only ${\cal N}=1$ supersymmetry.} versions of the flat superspace algebra involving additional bosonic and fermionic generators (some of them central) naturally arise.

The first enlarged supersymmetry algebra to be considered was introduced by Green \cite{Green:1989nn} in relation to the superstring ($p=1$), and incorporates a fermionic central charge $Z^{\a}$:
\bea
\, \{ Q_{\a}, Q_{\b} \} &=& -2 (C \G^a)_{\a \b} P_a \\
\, [ P_a,  Q_{\a}] &=&  \ii (C \G_a)_{\a \b} \, Z^{\b}.
\label{GA}
\eea
The standard Wess-Zumino term in the superstring action \cite{Green:1983sg, Henneaux:1984mh}, required to ensure kappa symmetry \cite{de Azcarraga:1982dw, de Azcarraga:1982xk, Siegel:1983hh}, is only invariant under supersymmetry transformations up to a total derivative. Siegel  \cite{Siegel:1994xr} showed that it is possible to define a manifestly supersymmetric Wess-Zumino terms using an enlarged superspace related to the ``Green algebra'' (\ref{GA}) by a coset construction.

Building on Siegel's paper, there is an extensive literature relating the embedding of $p$-branes and $D$-branes into superspace to enlarged supersymmetry algebras involving additional bosonic and fermionic generators beyond those appearing in the Green algebra.  These enlarged supersymmetry algebras have been motivated and derived by a variety of means, which fall into two broad categories. ``Top-down'' approaches have sought the most general enlarged superalgebras, using Maurer-Cartan equations and techniques based on free differential algebras \cite{DeAzcarraga:1989vh, Bergshoeff:1989ax, Townsend:1992fa, Bergshoeff:1992gq, Bergshoeff:1995hm, Sezgin:1996cj, Sakaguchi:1998kk, Chryssomalakos:1999xd, Deriglazov:1997cd}. ``Bottom-up'' approaches have explored modifications to the algebra of Noether currents for $p$-brane and $D$-brane actions due to quasi-invariant terms in the Lagrangian \cite{de Azcarraga:1989gm, DeAzcarraga:1991tm, Sorokin:1997ps, Hammer:1997ts, Bergshoeff:1998ha, Reimers:2005jf, Reimers:2005ar}.

In this paper, we provide a new and quite direct derivation for the enlarged supersymmetry algebra related to a given embedded world-volume structure, based on integrability of supersymmetry transformations. This generalises work already published in \cite{McArthur:2015pna}.

\section{Background}
\setcounter{equation}{0}

In the Green-Schwarz formulation \cite{Green:1983sg, Henneaux:1984mh}, $p$-branes are embeddings of a ($p + 1$)-dimensional bosonic world-volume into a superspace,
\be
\s ^i \rightarrow \left( x^a (\s), \q^{\a} (\s) \right),
\ee
where $\s^i$ are coordinates on the world-volume, and $(x^a , \q^{\a})$ are superspace coordinates. Here we consider flat  $D$-dimensional ${\cal N} = 1$ superspace. The supersymmetry algebra
\be
\{ Q_{\a}, Q_{\b} \} = - 2 \, (C \G^a)_{\a \b} \, P_a
\ee
is realised via the transformations of  superspace coordinates\footnote{Unless otherwise stated, spinors are Majorana, so $ \bar{\e} = \e^{T} C,$ where $C$ is the charge conjugation matrix. Also, the spacetime dimension must be such that $(C \G^a)_{\a \b}$ is symmetric. Exterior derivatives act to the right, so that if $\phi$ and $\omega$ are forms, $d (\phi \wedge \omega ) = d \phi \wedge \omega + (-1)^p \phi \wedge d \omega$ when $\phi$ is a $p$-form.}
\bea
\d_{\e} \, x^a &=& \ii \, (\bar{\e} \G^a \q) \\
\d_{\e} \, \q^{\a} &=&  \e^{\a}.
\label{SS}
\eea
In a group theoretic setting, these transformations result from the left group action on
\be
g (x, \theta) = e^{\ii(x^aP_a + \theta^{\a}Q_{\a})}.
\ee
The left-invariant one-forms $\pi^A, \, A = (a, \a)$, which are therefore invariant under supersymmetry transformations, are constructed as
\be
g(x, \theta)^{-1} d g(x, \theta) = \ii \p^{a}P_{a} + \ii \p^{\a} Q_{\a},
\ee
and are
\be
\p^{a} = dx^a - \ii (\bar{\q} \G^a d \q), \quad \p^{\a} = d \q^{\a}.
\ee
$p$-brane actions exist only in spacetime dimensions for which the supersymmetry invariant ($p+2$)-form
\be
h^{(p+2)} = \p^{a_1} \wedge \cdots \wedge \p^{a_p} \wedge (d \bar{\q} \wedge \G_{a_1} \cdots \G_{a_p} \, d \q)
\label{h}
\ee
is closed \cite{Bergshoeff:1987cm}, requiring the gamma matrix identities
\bea
0 & = & ( C\G^{a})_{\a ( \b} \, (C \G_{a} )_{\g \d )} , \quad p = 1 \label{id1} ; \label{p=1} \\
0 & = & ( C\G^{a_1})_{(\a \b} \, (C \G_{a_1 \cdots a_p})_{\g \d)},  \quad p > 1. \label{p>1}
\eea
Here, $ \G_{a_1 \cdots a_p}$ is the anti-symmetrized product of gamma matrices, and the round brackets on spinor indices denote symmetrisation. The resulting restrictions on $p$ and $D$  to ensure equal numbers of world-volume bosonic and fermionic degrees of freedom (and therefore world-volume supersymmetry) give  rise to the ``brane-scan'' \cite{Achucarro:1987nc, Evans:1988jb}.

The $p$-brane action is of the form
\be
S = S_0 + S_{WZ},
\label{action}
\ee
where the ``kinetic'' term
\be
S_0 = \int d^{(p+1)} \s \, \sqrt{ det \,G_{ij}}
 \ee
is constructed from the pull-back of the flat spacetime metric,
\be
G_{ij} = \p_i{}^a  \, \eta_{ab}  \,  \p_j{}^b,
\ee
with $\p_i{}^a = \frac{\partial x^a}{\partial \s^i} - \ii (\bar{\q} \G^a \frac{\partial \q}{\partial \s^i}).$
Since the de Rham cohomology of superspace is trivial, closure of the form $h^{(p+2)}$ in (\ref{h}) implies
 \be
h^{(p+2)} = d  b^{(p+1)}.
\ee
The Wess-Zumino term is given by the integral over the ($p$+1)-dimensional world-volume of the pullback of the superspace form $b^{(p+1)},$
\be
S_{WZ} = \int \s^* b^{(p+1)},
\label{WZ1}
\ee
and is necessary to ensure the $p$-brane action (\ref{action}) possesses a local fermionic $\k$-symmetry \cite{de Azcarraga:1982dw, de Azcarraga:1982xk, Siegel:1983hh}. This allows half of the world-volume fermionic degrees of freedom to be gauged away to ensure world-volume supersymmetry.
 The form $b^{(p+1)}$ cannot be chosen to be invariant under supersymmetry transformations (i.e. cannot be written in terms of the left-invariant one-forms $\pi^{A}$), but varies by a total derivative.
 Stated technically, $h^{(p+2)}$ belongs to a non-trivial  Chevalley-Eilenberg cohomology class on superspace, in which the cocycles are left-invariant one-forms  \cite{DeAzcarraga:1989vh}.

The ``quasi-invariance" of the Wess-Zumino Lagrangian under supersymmetry transformations (variation of the Lagrangian by a total derivative)  leads to modifications to the Noether currents generating supersymmetry transformations, which in turn leads to the corresponding charges obeying a modified algebra. In the original analysis \cite{de Azcarraga:1989gm}, the algebra was determined to be of the form
\be
\{ Q_{\a}, Q_{\b} \} = - 2 \, (C \G^a)_{\a \b} \, P_a + (C \G_{a_1 \cdots a_p})_{\a \b} \, Z^{a_1 \cdots a_p} ,
\ee
where $Z^{a_1 \cdots a_p} $ are bosonic central charges. This was based on solutions to the ``descent equations''
\be
h^{(p+2)} = d b^{(p+1)}, \quad \d_{\e} b^{(p+1)} = d \d_{\e} A^{(p)}
\ee
(where $\d_{\e}$ denotes a supersymmetry variation with parameter $\e^{\a}$, and the second equation follows from $\d_{\e} h^{(p+2)}= 0$) in which $b^{(p+1)}$ and
$\d_{\e} A^{(p)}$ were chosen to be invariant under spacetime translations. More general solutions which are not invariant under spacetime translations lead to additional  fermionic topological charges in the algebra of Noether charges \cite{Hammer:1997ts, Reimers:2005jf}. In the case $p=1$ (superstrings), the enlarged algebra is  the Green algebra \cite{Green:1989nn}.

The same enlarged supersymmetry algebras involving additional bosonic and fermionic generators have also emerged via an alternative route. As was pointed out by Siegel in the case of the superstring \cite{Siegel:1994xr}, and generalized to $p$-branes by others, it is possible to construct Wess-Zumino Lagrangians  that are manifestly invariant under spacetime supersymmetry transformations. The form $b^{(p+1)}$ in standard superspace with coordinates $(x, \theta)$ {\it cannot} be chosen to be supersymmetric (i.e. expressed in terms of left-invariant forms); however,  by introducing an enlarged superspace with additional coordinates, it is possible to introduce additional terms into the action to make it supersymmetric. The enlarged superspace is related by a coset construction \cite{Bergshoeff:1995hm,  Chryssomalakos:1999xd}  to  the enlarged  supersymmetry algebras that emerge via  ``quasi-invariance'' and modified Noether currents. Stated technically, $h^{(p+2)}$ belongs to a trivial Chevalley-Eilenberg cohomology class for the enlarged superspace, in that it can be expressed as $d(b^{(p+1)} - d A^{(p)} )$ where the additional superspace $p$-form $A^{(p)}$ is chosen so that $b^{(p+1)} - d A^{(p)}$ is invariant under supersymmetry transformations - and indeed can be expressed in terms of left-invariant forms on the enlarged superspace. The additional left-invariant forms  have been determined by a variety of means by different authors, including   using the concept of a free differential algebra \cite{DeAzcarraga:1989vh, Townsend:1992fa, Bergshoeff:1992gq, Chryssomalakos:1999xd}
 and requiring that the left-invariant  forms satisfy as consistent set of Maurer-Cartan equations (equivalent to the Jacobi identities of the corresponding algebra) \cite{Bergshoeff:1995hm, Sezgin:1996cj}.  The Jacobi identities rely on the gamma matrix identities (\ref{p=1}, \ref{p>1})  - the same identities required to ensure closure of the superspace forms (\ref{h}) giving rise to the Wess-Zumino terms.

The supersymmetry invariant Wess-Zumino term is then of the form
\be
S_{WZ} = \int  \s^* (b^{(p+1)} - d A^{(p)} ).
\label{WZ2}
\ee
This still satisfies the requirements of a Wess-Zumino term, namely that   $d (b^{(p+1)} - d A^{(p)}) $ is closed and supersymmetry invariant; the fact that the $p$-form $A^{(p)}$ appears as a total derivative means that the degrees of freedom it contains (including the additional superspace coordinates) do not contribute to the dynamics.

A similar issue is involved in the construction of $D$-brane actions.  As originally formulated, $D$-branes include a world-volume  $U(1)$ gauge field $A$ whose field strength $F = d A$ gives rise to a Born-Infeld term in the action \cite{Townsend:1995af, Aganagic:1996pe, Aganagic:1996nn, Cederwall:1996pv}; specifically,
\be
S =  \int d^{(p+1)} \s \, \sqrt{ \det \,(G_{ij}+ {\cal F}_{ij})} + S_{WZ}.
\ee
As with $p$-branes,  $G_{ij}$ is the pullback to the world-volume of the spacetime metric.  $ {\cal F} $ is the world-volume two-form ${\cal F} = F -  \s^*b^{(2)}, $ where $ d b^{(2)} = h^{(3)} $ is a supersymmetry invariant closed superspace three-form given by (\ref{h}) with $p = 1$. The Wess-Zumino term is again of the form
\be
S_{WZ} = \int \s^* b^{(p+1)},
\ee
with $h^{(p+2)} = db^{(p+1)}$ also supersymmetry invariant and closed as in (\ref{h}). $D$-branes are also required to possess  kappa symmetry, which again allows half of the fermionic degrees of freedom to be gauged away. The cohomology problem is essentially the same as that in the case of manifestly supersymmetric Wess-Zumino terms for $p$-branes: to find a world-volume one-form $A$ such that ${\cal F} = dA-  \s^*b^{(2)} $ is invariant under spacetime supersymmetry transformations.

Similarly, the M theory 5-brane action involves a  world-volume two-form gauge field $A^{(2)}$ such that
$ d A^{(2)} - \sigma^* b^{(3)} $ is  spacetime supersymmetry invariant \cite{Aganagic:1997zq}.

In this paper, we consider an alternative and more direct approach to the determination of the enlarged supersymmetry algebras associated with a given $p$-brane  or $D$-brane action. We consider the integrability of the supersymmetry transformations $\d_{\e} A^{(p)}$ assigned to $A^{(p)}$ in solving the cohomology problem
\be
h^{(p+2)} = d b^{(p+1)}, \quad \d_{\e} b^{(p+1)} = d \d_{\e} A^{(p)}.
\ee
Not only does this lead us directly to the enlarged supersymmetry algebra associated with a given brane; it also leads us to the representation in standard superspace  of bosonic and fermionic Noether charges associated with the enlarged algebra. This generalises earlier results reported in the case of $p=1$ \cite{McArthur:2015pna}.

\section{Review of the case $p$ = 1}
\setcounter{equation}{0}

This case was examined in \cite{McArthur:2015pna}, and relates to both the Wess-Zumino term in the action for the superstring, and the worldvolume one-form appearing in the Born-Infeld term in the action for $D$-branes. We begin with the supersymmetry invariant and closed three-form\footnote{From here onward we do not include the symbol $\wedge$ in the wedge product of forms.}
\be
 h^{(3)} = \p^a  (d \bar{\q} \G_a d \q).
 \ee
The most general candidate for $b^{(2)}$ is
\be
b^{(2)} = \m \,x^a (d \bar{\q} \G_a d \q) - (1 - \m) \, dx^a(\bar{\q} \G_a  d \q),
\label{b2}
\ee
where $\m$ is a real parameter. Varying $\m$ changes $b^{(2)} $ by an exterior derivative, so it is an ``integration constant'' in solving  $h^{(3)} = d b^{(2)}.$ Conventionally $\m$ is set to zero in the literature to make $b^{(2)}$ invariant under spacetime translations.  Requiring
$ \d_{\e} b^{(2)} = d \d_{\e} A^{(1)} ,$ we obtain
\be
\d_{\e} A^{(1)} = (1- \m) \n \, dx^a (\bar{\e} \G_a \q) - \frac{\ii}{3} (1 - 3 \m) \, (\bar{\e} \G^a \q) (\bar{\q} \G_a d \q) - (1 - \m)(1 - \n) \, x^a (\bar{\e} \G_a d \q),
\label{deA}
\ee
where again $\n$ is an integration constant, and again conventionally set to 1 to make $\d_{\e} A^{(1)} $ spacetime translation invariant.

We can check the integrability of the supersymmetry transformation by computing the commutator of two supersymmetry transformations:
\be
(\d_{\e_2} \d_{\e_1} - \d_{\e_1} \d_{\e_2} ) A^{(1)} = 2 (\bar{\e_1} \G_a \e_2 )\, \left( (1 - \m) \n \, dx^a + \ii \m \, (\bar{\q} \G^a d \q) \right).
\label{ddA}
\ee
If we use $\d_{\e} =  \e^{\a} Q_{\a} ,$ and require consistency of (\ref{ddA}) with the anticommutator $\{ Q_{\a}, Q_{\b} \} = -2 (C \G^a)_{\a \b} P_a,$
then we infer that as an operator relation in ``$(x, \q, A)$ space"\footnote{Compatibility of $(\d_{\e_2} \d_{\e_1} - \d_{\e_1} \d_{\e_2} ) x^a = 2 \ii (\bar{\e_1} \G^a \e_2 ) $ with the algebra requires $P_a x^b = - \ii \d_a{}^b$.},
\be
P_a A^{(1)} = - (1 - \m) \n dx_a - \ii \m (\bar{\q} \G_a d \q).
\label{PaA}
\ee
It then follows that
\be
[ P_a ,Q_{\a} ] A^{(1)} = \ii (C \G_a  d\q )_{\a},
\label{PQA}
\ee
which is not consistent with the the standard supersymmetry algebra which has $[ P_a ,Q_{\a} ] = 0.$ Instead, we have a realisation of an extension of the standard supersymmetry algebra by an additional fermionic charge.
This is consistent with the Green algebra (\ref{GA}) if the action of the charge $Z^{\a}$ is realized  via
\be
Z^{\a} A^{(1)} =  d\q ^{\a}.
\ee
It is easy to check that $ Z^{\a} $ is a central charge if we assume the superspace coordinates $(x, \theta)$ are inert under the action of $ Z^{\a}. $ Note that the resulting algebra is independent of the choice of the integration constants $\m$ and $\n$.

Thus we naturally see the emergence of the Green algebra, already known to be applicable in formulating manifestly supersymmetric Wess-Zumino terms for  the superstring \cite{Siegel:1994xr}, in the solution of $ \d_{\e} b^{(2)} = d \d_{\e} A^{(1)}.$

\section{The case $p=2$}\label{p=2a}
\setcounter{equation}{0}

We now consider the case where $p=2$, which relates to the construction of a supersymmetric Lagrangian for a 2-brane (or super-membrane), and to the two-form gauge field on the worldvolume of an  M5-brane \cite{Aganagic:1997zq}.  In this case the four-form
\begin{equation}
  h^{(4)}=\p^{a}\p^{b}(d\ba{\q}\G_{ab}d\q)
\end{equation}
is closed provided
\begin{equation}
 0=(C\G^{a})_{(\a\b}(C \G_{ab})_{\g\d)}\,. \label{gammaid3}
\end{equation}
Since $h^{(4)}$ is closed,
\begin{equation}
  h^{(4)}=d b^{(3)}\,.
\end{equation}
The most general expression (including terms not invariant under spacetime translations) for the three-form $b^{(3)}$ is
\begin{multline}
  b^{(3)}=(1-\a) x^a dx^b(d\ba{\q}\G_{ab}d\q)+\a dx^a dx^b(\ba{\q}\G_{ab}d\q)\\-\ii dx^a(\ba{\q}\G^bd\q)(\ba{\q}\G_{ab}d\q)-\frac13(\ba{\q}\G^ad\q)(\ba{\q}\G^bd\q)(\ba{\q}\G_{ab}d\q)+d c^{(2)}\label{b3}
\end{multline}
where $\a$ is a constant and $c^{(2)}$ is any two-form.  In deriving this result we have used the identity
\begin{align}
  (\ba{\q}\G^a d\q)(d \ba{\q}\G_{ab}d\q)= - (d\ba{\q}\G^a d\q)(\ba{\q}\G_{ab}d\q)  \label{gammaid4}
\end{align}
which follows from \eqref{gammaid3}.

Since $h^{(4)}=d b^{(3)}$ is invariant under supersymmetry transformations we have
\begin{equation}
  \d_\e b^{(3)}=d a^{(2)}(\e)\,,
  \label{deltab3}
\end{equation}
where $\d_\e$ denotes a supersymmetry transformation with parameter $\e^\a$, and $a^{(2)}(\e)$ is some $\e$ dependent two-form.  The most general expression for $a^{(2)}(\e)$ is
\begin{align}
   a^{(2)}(\e)&= \a(1-\b) x^a dx^b(\ba{\e}\G_{ab}d\q)+\a\b dx^a dx^b(\ba{\e}\G_{ab}\q)\no\\
   &+\ii (1-\a)(1-\g)x^a (\ba{\e}\G^{b}\q)(d\ba{\q}\G_{ab}d\q)-\ii (1-\a)\g x^a (\ba{\e}\G^{b}d\q)(\ba{\q}\G_{ab}d\q)\no\\&
   -\ii (2 \a+\g-\a\g-\tfrac53)dx^a (\ba{\e}\G^{b}\q)(\ba{\q}\G_{ab}d\q)-\frac{\ii}{3}dx^a (\ba{\q}\G^{b}d\q)(\ba{\e}\G_{ab}\q)\no\\
   &+\frac{1}{15}(\ba{\e}\G^{a}\q)(\ba{\q}\G^{b}d\q)(\ba{\q}\G_{ab}d\q)-\frac{1}{15}(\ba{\q}\G^{a}d\q)(\ba{\q}\G^{b}d\q)(\ba{\e}\G_{ab}\q)+\d_\e c^{(2)}+d a^{(1)}(\e)
\end{align}
where $\a$, $\b$, and $\g$ are all constants and $a^{(1)}(\e)$ is any $\e$ dependent one-form.  In deriving this result we have exploited the identity \eqref{gammaid3} a number of times; for example  \eqref{gammaid4} and
\begin{align}
  (d\ba{\q}\G^a d\q)(\ba{\e}\G_{ab}\q)+(\ba{\e}\G^a \q)(d\ba{\q}\G_{ab}d\q)= 2(\ba{\e}\G^a d\q)(\ba{\q}\G_{ab}d\q)-2(\ba{\q}\G^a d\q)(\ba{\e}\G_{ab}d\q)  \label{gammaid5}
\end{align}
have been used.

We now introduce a two-from $A^{(2)}$ whose supersymmetry transformation is determined by the requirement that $\mathcal{F}^{(3)}=d A^{(2)}-b^{(3)}$ is invariant.  Without loss of generality this implies that
\begin{equation}
  \d_\e A^{(2)}= a^{(2)}(\e)\,,
\end{equation}
and so we arrive at the  supersymmetry transformation required of the two-form $A^{(2)}$:
\begin{align}
  \d_\e A^{(2)}&= \a(1-\b) x^a dx^b(\ba{\e}\G_{ab}d\q)+\a\b dx^a dx^b(\ba{\e}\G_{ab}\q)\no\\
   &+\ii (1-\a)(1-\g)x^a (\ba{\e}\G^{b}\q)(d\ba{\q}\G_{ab}d\q)-\ii (1-\a)\g x^a (\ba{\e}\G^{b}d\q)(\ba{\q}\G_{ab}d\q)\no\\&
   -\ii (2 \a+\g-\a\g-\tfrac53)dx^a (\ba{\e}\G^{b}\q)(\ba{\q}\G_{ab}d\q)-\frac{\ii}{3}dx^a (\ba{\q}\G^{b}d\q)(\ba{\e}\G_{ab}\q)\no\\
   &+\frac{1}{15}(\ba{\e}\G^{a}\q)(\ba{\q}\G^{b}d\q)(\ba{\q}\G_{ab}d\q)-\frac{1}{15}(\ba{\q}\G^{a}d\q)(\ba{\q}\G^{b}d\q)(\ba{\e}\G_{ab}\q)+\d_\e c^{(2)}+d a^{(1)}(\e)\,.
   \label{ssva}
\end{align}

Up to redefinitions of the constants $\a$, $\b$ and $\g$, the most general choices for $c^{(2)}$ and $a^{(1)}(\e)$  in $\d_\e A^{(2)}$ are
\begin{align}
  c^{(2)}&= \ii \l x^a (\ba{\q}\G^{b}d\q)(\ba{\q}\G_{ab}d\q) \label{c2}
\intertext{and}
  a^{(1)}(\e)&= \ii \k x^a (\ba{\q}\G^{b}d\q)(\ba{\e}\G_{ab}\q) \label{a1}
\end{align}
with $\k$ and $\l$ constants.

We can now check the integrability of the supersymmetry transformation (\ref{ssva}) assigned to $A^{(2)}$ by computing the commutator of two supersymmetry transformations. As detailed in Appendix \ref{appendix}, the resulting expression for the commutator can be expressed via terms of the form $(\ba{\e}_1\G^{a}\e_2)$ and $(\ba{\e}_1\G_{ab}\e_2)$ -- where the former relates to the standard supersymmetry algebra, and the latter can be interpreted via an additional charge in the algebra (see below) --  provided we set
\begin{align}
  \k=\frac{1}{15}\,,\qquad \b=\frac{1}{3\a}\,,\qquad  \g=1 + \frac{1}{15(\a-1)}\,, \label{consts}
\end{align}
with the other two constants, $\a$ and $\l$, remaining unfixed.  Our result is
\begin{align}
  [\d_{\e_2},\d_{\e_1}]A^{(2)}=\,&\,2\ii(\ba{\e}_1\G^{a}\e_2)\left\{\frac{1}{15} x^b(d\ba{\q} \G_{ab}d\q)+(\a-\tfrac{14}{15})d x^b(\ba{\q} \G_{ab}d\q)+\ii \l(\ba{\q}\G^{b}d\q)(\ba{\q}\G_{ab}d\q)\right\}\no\\
  &+\frac23(\ba{\e}_1\G_{ab}\e_2)\left\{dx^a d x^b-\frac{\ii}{5}d\big[x^a(\ba{\q} \G^{b}d\q)\big]\right\}\,.\label{d12Aa}
\end{align}
If we use $\d_\e=\e^\a Q_\a$ and require that \eqref{d12Aa} be consistent with an extension of the supersymmetry algebra of the form
\begin{equation}
  \left\{Q_\a,Q_\b\right\}=-2(C\G^a)_{\a\b}P_a+ (C\G_{ab})_{\a\b}Z^{{ab}}
\end{equation}
where $Z^{ab}=-Z^{ba}$ is a bosonic  charge, then we infer that as an operator relation in ``$(x,\q,A)$ space'',
\begin{align}
  P_a A^{(2)}&=-\frac{\ii}{15} x^b(d\ba{\q} \G_{ab}d\q)-\ii(\a-\tfrac{14}{15})d x^b(\ba{\q} \G_{ab}d\q)+ \l(\ba{\q}\G^{b}d\q)(\ba{\q}\G_{ab}d\q) \label{PA}
  \end{align}
and\footnote{Note that our (anti)symmetrization of $n$ indices carries a factor of $1/n!$.}
\begin{align}
 Z^{ab}A^{(2)} &=d\left[\frac23 x^a d x^b-\frac{2\ii}{15}x^{[a}(\ba{\q} \G^{b]}d\q)\right]\,.\label{ZabA}
\end{align}

\section{Completing the $p=2$ algebra}\label{p=2b}
\setcounter{equation}{0}

We are now in a position to determine the rest of the algebra.  For example, from \eqref{ssva} with the choices \eqref{c2}, \eqref{a1} and \eqref{consts}, and by computing the supersymmetric variation of \eqref{PA} and using $\d_\e=\e^\a Q_\a$ we find
\begin{align}
[Q_\a,P_a]A^{(2)}=d\left[ \frac{3  \ii}{5} x^b (d\ba{\q}\G_{ab})_\a-\frac{1}{15}(\ba{\q}\G^{b}d\q)(\ba{\q}\G_{ab})_\a\right]
\end{align}
which leads us to introduce a fermionic charge $Z^{a\a}$ defined via
\begin{align}
[Q_\a,P_a]=(C\G_{ab})_{\a\b}Z^{b\b}
\end{align}
where
\begin{align}
 Z^{a\a}A^{(2)} &=d\left[ \frac{3  \ii}{5} x^a d\q^\a-\frac{1}{15}(\ba{\q}\G^{a}d\q)\q^\a\right]\,.
\end{align}

Similarly, computing the supersymmetry variation of \eqref{ZabA} and using $\d_\e=\e^\a Q_\a$, and assuming that $x^a$ and $\q^\a$ are inert under the action of the additional charges we find
\begin{align}
 [Q_\a,Z^{ab}]A^{(2)} &=d\left[ \frac{6  \ii}{5} x^{[a} (d\ba{\q} \G^{b]})_\a-\frac{2}{15}   (\ba{\q} \G^{[a} d\q)(\ba{\q} \G^{b]})_\a\right]\no\\
 &=-2 (C\G^{[a})_{\a\b}Z^{b]\b}A^{(2)}\,.
\end{align}

In this way we can construct the entire algebra, which is
\begin{align}
  \left\{Q_\a,Q_\b\right\}&=-2(C\G^a)_{\a\b}P_a+ (C\G_{ab})_{\a\b}Z^{{ab}}   \label{p=2algebra1}\\
  [Q_\a,P_a]&=(C\G_{ab})_{\a\b}Z^{b\b}\\
  [P_a,P_b]&=(C\G_{ab})_{\a\b}Z^{\a\b} \\
  [Q_\a,Z^{ab}]&=-2 (C\G^{[a})_{\a\b}Z^{b]\b} \\
  [P_c,Z^{ab}]&=- \d_c^{[a}(C\G^{b]})_{\a\b}Z^{\a\b} \\
  \left\{Q_\a,Z^{a\b}\right\}&=-4 (C\G^{a})_{\a\g}Z^{\g\b}-\frac12 \d_{\a}^{\b} (C\G^{a})_{\g\l}Z^{\g\l},
  \label{p=2algebra2}
\end{align}
where
\begin{align}
 P_a A^{(2)}&=-\frac{\ii}{15} x^b(d\ba{\q} \G_{ab}d\q)-\ii(\a-\tfrac{14}{15})d x^b(\ba{\q} \G_{ab}d\q)+ \l(\ba{\q}\G^{b}d\q)(\ba{\q}\G_{ab} d\q) \label{p2charges1} \\
 Z^{ab}A^{(2)} &=d\left[\frac23 x^a d x^b-\frac{2\ii}{15}x^{[a}(\ba{\q} \G^{b]}d\q)\right]  \\
 Z^{a\a}A^{(2)} &=d\left[ \frac{3\ii}{5} x^a d\q^\a-\frac{1}{15}(\ba{\q}\G^{a}d\q)\q^\a\right]\\
 Z^{\a\b}A^{(2)} &=\frac{2}{15}d\left[\q^{\a} d\q^{\b}\right].
 \label{p2charges2}
\end{align}
Note that the algebra (\ref{p=2algebra1})-(\ref{p=2algebra2})  is independent of the choice of the constants $\alpha$ and $\lambda.$ Up to redefinitions of the extra generators, it is equivalent to the $p=2$  algebra identified by Bergshoeff and Sezgin  \cite{Bergshoeff:1995hm} in their construction of a $p=2$ supersymmetric Wess-Zumino term via an expanded superspace. The difference is that we have systematically derived the algebra; in  \cite{Bergshoeff:1995hm}, the algebra is stated without derivation. The right hand sides of (\ref{p2charges1})-(\ref{p2charges2}) also reproduce the Noether charges constructed in \cite{Reimers:2005cp} (up to a different choice of  normalization).

\section{$p=3$ algebra}
\setcounter{equation}{0}

Utilising the same procedure as that used for $p=2$, we now extend the derivation of the enlarged algebra to the case of $p=3$.  Here we  summarise our results.

Starting with $h^{(5)}=d b^{(4)}$, we find that the most general expression for the four-form $b^{(4)}$ is given by
\begin{align}
  b^{(4)}=\,&\,(1+m_1) x^a dx^b dx^c(d\ba{\q}\G_{abc}d\q)
  +m_1 dx^a dx^b dx^c(\ba{\q}\G_{abc}d\q) \no \\
  &+2 m_2 x^a dx^b(\ba{\q}\G^c d\q)(d\ba{\q}\G_{abc}d\q)
  +\left(\tfrac{3 \ii}{2}+m_2\right) dx^a dx^b(\ba{\q}\G^c d\q)(\ba{\q}\G_{abc}d\q) \no\\
  & +3 m_3 x^a(\ba{\q}\G^b d\q)(\ba{\q}\G^c d\q)(d\ba{\q}\G_{abc}d\q)
  +(1+m_3)dx^a(\ba{\q}\G^b d\q)(\ba{\q}\G^c d\q)(\ba{\q}\G_{abc}d\q)\no \\
   &-\frac{\ii}{4}(\ba{\q}\G^a d\q)(\ba{\q}\G^b d\q)(\ba{\q}\G^c d\q)(\ba{\q}\G_{abc}d\q)
\end{align}
where $m_1$, $m_2$ and $m_3$ are constants.

Requiring
\begin{equation}
  \d_\e b^{(4)}=d \d_\e A^{(3)},
\end{equation}
we find that
\begin{align}
\d_\e A^{(3)}=\,& (m_1+m_4) x^a dx^b dx^c(\ba{\e}\G_{abc}d\q)+m_4 dx^a dx^b dx^c(\ba{\e}\G_{abc}\q)\no\\
  &+(\ii+\ii m_1-m_5+m_6) x^a dx^b (\ba{\e}\G^{c}\q)(d\ba{\q}\G_{abc}d\q) \no \\
  &+(\tfrac{3\ii}{2}+3 \ii m_1+m_2-m_5-2m_6) x^a d x^b (\ba{\e}\G^{c}d\q)(\ba{\q}\G_{abc}d\q)\no\\
  &+(\tfrac{3\ii}{2}+m_2+3m_6) x^a d x^b (\ba{\q}\G^{c}d\q)(\ba{\e}\G_{abc}d\q)\no\\
  &+ m_5 d x^a d x^b (\ba{\e}\G^{c}\q)(\ba{\q}\G_{abc}d\q)+ m_6 d x^a d x^b (\ba{\q}\G^{c}d\q)(\ba{\e}\G_{abc}\q)\no\\
  &+ (2\ii m_2+2 m_3+m_7-4m_8-1)   x^a(\ba{\e}\G^{b}d\q) (\ba{\q}\G^{c}d\q)(\ba{\q}\G_{abc}d\q)\no \\
  &-2 (\ii m_2+m_7+m_8)   x^a(\ba{\e}\G^{b}\q) (\ba{\q}\G^{c}d\q)(d\ba{\q}\G_{abc}d\q) \no \\
  &+ (1+ m_3+5m_8)   x^a(\ba{\q}\G^{b}d\q) (\ba{\q}\G^{c}d\q)(\ba{\e}\G_{abc}d\q)\no \\
  &+ m_7   d x^a(\ba{\e}\G^{b}\q) (\ba{\q}\G^{c}d\q)(\ba{\q}\G_{abc}d\q)+ m_8   d x^a(\ba{\q}\G^{b}d\q) (\ba{\q}\G^{c}d\q)(\ba{\e}\G_{abc}\q)\no \\
  &+ \frac{\ii}{28}(1+28 m_3) (\ba{\e}\G^{a}\q) (\ba{\q}\G^{b}d\q) (\ba{\q}\G^{c}d\q)(\ba{\q}\G_{abc}d\q)\no \\
  &+ \frac{\ii}{28} (\ba{\q}\G^{a}d\q) (\ba{\q}\G^{b}d\q) (\ba{\q}\G^{c}d\q)(\ba{\e}\G_{abc}\q) \label{ssvarA3}
\end{align}
where $m_i$, with $i=1,2, \ldots, 8$, are all constants.

We can now check the integrability of the supersymmetry transformation assigned to $A^{(3)}$ by computing the commutator of  two supersymmetry transformations. The resulting expression for the commutator can be expressed via terms of the form $(\ba{\e}_1\G^{a}\e_2)$ and $(\ba{\e}_1\G_{abc}\e_2)$ -- the latter being able to be interpreted in terms of an additional charge in the algebra  --  provided we set
\begin{align}
  m_4=\frac{1}{4}\,,\quad m_5=\frac{93 \,  \ii}{140}+\ii  \,m_1\,,\quad  m_6=- \frac{47 \, \ii }{140}\,,\quad  m_7= \frac{5}{28}-\ii \, m_2\,, \quad  m_8= -\frac{5}{28}\,,
\end{align}
with $m_1$, $m_2$ and $m_3$ remaining unfixed.  Our result is
\begin{multline}
  [\d_{\e_2},\d_{\e_1}]A^{(3)}= - 2(\ba{\e}_1\G^{a}\e_2)\bigg\{
  -\ii \Big((m_1+\tfrac{61}{70}) dx^b dx^c +(m_2+\tfrac{3\ii}{140}) dx^b (\ba{\q}\G^{c} d\q)  \\  + m_3 (\ba{\q}\G^{b} d\q)(\ba{\q}\G^{c} d\q) \Big)(\ba{\q}\G_{abc} d\q)-\tfrac{3}{70}x^b\Big(3 \ii dx^c+(\ba{\q}\G^{c} d\q) \Big)(d\ba{\q}\G_{abc} d\q)
  \bigg\}\\
  +(\ba{\e}_1\G_{abc}\e_2)d\left[\frac12 x^a dx^b dx^c-\frac{9\ii}{35}x^{[a} dx^b(\ba{\q}\G^{c]} d\q)-\frac{3}{70}x^{[a} (\ba{\q}\G^{b} d\q)(\ba{\q}\G^{c]} d\q)\right]\,.
\end{multline}
If we use $\d_\e=\e^\a Q_\a$ and require that the above expression be consistent with an extension of the supersymmetry algebra of the form
\begin{equation}
  \left\{Q_\a,Q_\b\right\}=-2(C\G^a)_{\a\b}P_a+ (C\G_{abc})_{\a\b}Z^{{abc}}
\end{equation}
where $Z^{abc}$ is a bosonic charge completely antisymmetric in its indices, then we infer that as an operator relation in ``$(x,\q,A)$ space'',
\begin{align}
 P_a A^{(3)}=&-\ii \Big((m_1+\tfrac{61}{70}) dx^b dx^c +(m_2+\tfrac{3\ii}{140}) dx^b (\ba{\q}\G^{c} d\q) + m_3 (\ba{\q}\G^{b} d\q)(\ba{\q}\G^{c} d\q) \Big)(\ba{\q}\G_{abc} d\q)\no \\&-\tfrac{3}{70}x^b\Big(3 \ii dx^c+(\ba{\q}\G^{c} d\q) \Big)(d\ba{\q}\G_{abc} d\q)
\end{align}
and
\begin{align}
Z^{abc}A^{(3)}=d\left[\frac12 x^a dx^b dx^c-\frac{9\ii}{35}x^{[a} dx^b(\ba{\q}\G^{c]} d\q)-\frac{3}{70}x^{[a} (\ba{\q}\G^{b} d\q)(\ba{\q}\G^{c]} d\q)\right]
\end{align}

Continuing as in the $p=1$ and $p=2$ cases, we find the resulting algebra is
\begin{align}
  \left\{Q_\a,Q_\b\right\}&=-2(C\G^a)_{\a\b}P_a+ (C\G_{abc})_{\a\b}Z^{{abc}} \label{p=3algebra1}\\
  [Q_\a,P_a]&=(C\G_{abc})_{\a\b}Z^{bc\b}\\
  [P_a,P_b]&=(C\G_{abc})_{\a\b}Z^{c\a\b} \\
  [Q_\a,Z^{abc}]&=-2 (C\G^{[a})_{\a\b}Z^{bc]\b} \\
  [P_e,Z^{abc}]&=- \d_e^{[a}(C\G^{b})_{\a\b}Z^{c]\a\b} \\
  [P_a,Z^{bc\b}]&=-\d_a^{[b}(C\G^{c]})_{\a\g}Z^{\a\g\b} \\
 \left\{Q_\a,Z^{ab\b}\right\}&=-4 (C\G^{[a})_{\a\g}Z^{b]\g\b}-\frac12 \d_{\a}^{\b} (C\G^{[a})_{\g\l}Z^{b]\g\l}\\
  [Q_\a,Z^{a\b\g}]&=-\frac52 (C\G^{a})_{\a\l}Z^{\l\b\g}-\frac12  (C\G^{a})_{\l\m}Z^{\l\m(\g}\d_{\a}^{\b)}
  \label{p=3algebra2}
\end{align}
with
\begin{align}
 P_a A^{(3)}=&-\ii \Big((m_1+\tfrac{61}{70}) dx^b dx^c +(m_2+\tfrac{3\ii}{140}) dx^b (\ba{\q}\G^{c} d\q) + m_3 (\ba{\q}\G^{b} d\q)(\ba{\q}\G^{c} d\q) \Big)(\ba{\q}\G_{abc} d\q)\no \\&-\tfrac{3}{70}x^b\Big(3 \ii dx^c+(\ba{\q}\G^{c} d\q) \Big)(d\ba{\q}\G_{abc} d\q) \label{PA3}
\end{align}
with $m_1$, $m_2$ and $m_3$ arbitrary constants, and
\begin{align}
Z^{abc}A^{(3)}&=d\left[\frac12 x^a dx^b dx^c-\frac{9 \, \ii}{35}x^{[a} dx^b(\ba{\q}\G^{c]} d\q)-\frac{3}{70}x^{[a} (\ba{\q}\G^{b} d\q)(\ba{\q}\G^{c]} d\q)\right]\\
Z^{ab\a}A^{(3)}&=-d\left[\frac{\ii}{4}dx^a dx^b \q^\a+\frac{13 \, \ii}{35} x^a dx^b d\q^\a+\frac{3}{35}x^a(\ba{\q}\G^{b} d\q)d\q^\a\right.
\no \\ &\,\,\left. \qquad \qquad \qquad +\frac{9}{70}dx^a(\ba{\q}\G^{b} d\q)\q^\a -\frac{3 \, \ii}{140}(\ba{\q}\G^{a} d\q)(\ba{\q}\G^{b} d\q)\q^\a\right]\\
Z^{a\a\b}A^{(3)}&=d\left[\frac{1}{14}x^a d\q^\a d\q^\b-\frac{13}{70} dx^a \q^{(\a}d\q^{\b)}-\frac{3 \, \ii}{70}(\ba{\q}\G^{a} d\q)\q^{(\a}d\q^{\b)}\right]\\
Z^{\a\b\g}A^{(3)}&=-\frac{3\, \ii}{35}d\q^\a d\q^\b d\q^\g\,.
\label{p3charges}
\end{align}
The algebra is independent of the choice of the constants $m_1$, $m_2$ and $m_3$, and is equivalent to the $p=3$  algebra identified in \cite{Bergshoeff:1995hm}, which was stated without derivation; the algebra has been derived systematically in the approach we are using.   The right hand side of (\ref{p3charges}) again reproduces the $p=3$ Noether charges constructed in \cite{Reimers:2005cp} (up to normalization).

\section{$p=3$ enlarged superspace}
\setcounter{equation}{0}
Manifestly supersymmetric Wess-Zumino terms for $p$-branes  can be constructed as
\be
S_{WZ} = \int  \s^* (b^{(p+1)} - d A^{(p)} ),
\ee
where the supersymmetry transformations of the form $A^{(p)}$ are determined from those of $b^{(p+1)}.$ As shown in this paper, the integrability of the spacetime supersymmetry transformations assigned to $A^{(p)}$ reveals an underlying enlarged supersymmetry algebra, generalizing the Green algebra from the case $p=1.$ As shown by Siegel in the latter case \cite{Siegel:1994xr}, $A^{(1)}$ can be constructed as a form on an enlarged superspace related to the Green algebra by a coset construction. This was extended to the case $p=2$ in \cite{Bergshoeff:1995hm} and \cite{Chryssomalakos:1999xd}.

Here, we explicitly construct $A^{(3)}$ as a form in the enlarged superspace associated with the algebra \eqref{p=3algebra1}-\eqref{p=3algebra2}.  The superspace is defined in terms of the coset
\begin{multline}
  \W(x, \q,y,\c,w, \f)\\=\exp\left\{\ii (x^a P_a+\q^\a Q_\a+y_{abc}Z^{abc}+\c_{ab\a}Z^{ab\a}+w_{a\a\b}Z^{a\a\b}+\f_{\a\b\g}Z^{\a\b\g} )\right\}\,,
  \label{Omega}
\end{multline}
where  the isotropy subgroup is the Lorentz group. Transformations of the enlarged superspace coordinates $(x, \q, y, \c , w, \f)$ are then determined by considering the left action of group elements $g$ on  $\W(x, \q, y, \c , w, \f)$.  For instance, a supersymmetry transformation with parameter $\e^{\a}$ is achieved by the left action of the group element
\begin{align}
  g=\ex^{\ii \e^{\a}Q_{\a}}
\end{align}
on $\W(x, \q, y, \c , w, \f)$.  For an infinitesimal supersymmetry transformation we find the enlarged superspace coordinates transform as follows:
\begin{align}
  \d_\e x^a = \,&\,\ii (\ba{\e}\G^a\q) \label{ssp31}\\
  \d_\e \q^\a= \,&\,\e^\a\\
  \d_\e y_{abc}= \,&\,-\frac{\ii}{2} (\ba{\e}\G_{abc}\q) \\
\d_\e \c_{ab\a}= \,&\,\frac{\ii}{2}x^c (\ba{\e}\G_{cab})_{\a}-\ii y_{abc} (\ba{\e}\G^{c})_{\a}
   +\frac{1}{6}(\ba{\e}\G^{c}\q)(\ba{\q}\G_{cab})_{\a}+\frac{1}{6}(\ba{\e}\G_{cab}\q)(\ba{\q}\G^{c})_{\a} \\
\d_\e w_{a\a\b}=\,&\,2 \ii \c_{ab(\a}(\ba{\e}\G^b)_{\b)}-\frac{\ii}{4}\e^\g\c_{ab\g} (C\G^b)_{\a\b}+\frac{1}{24}x^b(\ba{\e}\G_{bca}\q)(C\G^c)_{\a\b}\no \\\,&\,-\frac13 x^b(\ba{\e}\G_{bca})_{(\a}(\ba{\q}\G^c)_{\b)}+\frac16 x^b(\ba{\e}\G^{c}\q)(C\G_{bca})_{\a\b}+\frac14 y_{abc}(\ba{\e}\G^{b}\q)(C\G^{c})_{\a\b}\no \\ \,&\,+\frac23 y_{abc}(\ba{\e}\G^{b})_{(\a}(\ba{\q}\G^c)_{\b)}\\
\d_\e \f_{\a\b\g}=\,&\,\frac56\c_{ab(\a}(\ba{\q}\G^{a})_{\b}(\ba{\e}\G^{b})_{\g)}-\frac14\c_{ab(\a}(C\G^{a})_{\b\g)}(\ba{\e}\G^{b}\q)
-\frac{1}{12}\q^\l\c_{ab\l}(C\G^{a})_{(\a\b}(\ba{\e}\G^{b})_{\g)}\no\\ &-\frac{1}{12}\e^\l\c_{ab\l}(C\G^{a})_{(\a\b}(\ba{\q}\G^{b})_{\g)}
-\frac{5\ii}{4}w_{a(\a\b}(\ba{\e}\G^{a})_{\g)}-\frac{\ii}{4}\e^\l w_{a\l(\a}(C\G^{a})_{\b\g)}\no\\
&+\frac16 x^ay_{abc}(C\G^{b})_{(\a\b}(\ba{\e}\G^{c})_{\g)}
+\frac{1}{36}(\ba{\q}\G^{a})_{(\a}(\ba{\q}\G^{b})_{\b}(\ba{\q}\G_{abc})_{\g)}(\ba{\e}\G^{c}\q)\no\\
&+\frac{1}{36}(\ba{\q}\G^{a})_{(\a}(\ba{\q}\G^{b})_{\b}(\ba{\q}\G^{c})_{\g)}(\ba{\e}\G_{abc}\q)\,. \label{ssp32}
\end{align}

An explicit expression for $A^{(3)}$ in terms of the enlarged superspace coordinates can now be constructed.   Given the known effect of the generators $P_a$, $Z^{abc}$, $Z^{ab\a}$, $Z^{a\a\b}$ and $Z^{\a\b\g}$ on $A^{(3)}$, \eqref{PA3}-\eqref{p3charges}, and the infinitesimal transformations of the enlarged superspace coordinates under the action of these generators (such as \eqref{ssp31}-\eqref{ssp32}), we can infer the types of terms which  must appear in $A^{(3)}$.  The coefficients of these terms can then be fixed by insisting that equation \eqref{ssvarA3} holds.  Our final result is:

\begin{align}
  A^{(3)}=\, &\,\left(m_1+\tfrac{157}{280}\right)x^a d x^b d x^c (\ba{\q}\G_{abc} d\q)+\left(m_2+\tfrac{3\ii}{10}\right)x^a d x^b(\ba{\q}\G^{c} d\q)(\ba{\q}\G_{abc} d\q)\no\\& +\left(m_3+\tfrac{3}{40}\right)x^a (\ba{\q}\G^{b} d\q)(\ba{\q}\G^{c} d\q)(\ba{\q}\G_{abc} d\q)
  + \frac{\ii}{2} y_{abc}d x^a d x^b d x^c+\frac{123}{140} y_{abc}d x^a d x^b (\ba{\q}\G^{c} d\q)\no \\ &
  -\frac{3 \ii}{5} y_{abc}d x^a (\ba{\q}\G^{b} d\q)(\ba{\q}\G^{c} d\q)-\frac{3}{20} y_{abc}(\ba{\q}\G^{a} d\q)(\ba{\q}\G^{b} d\q)(\ba{\q}\G^{c} d\q)
  \no  \\& -\frac{9}{70} y_{abc}x^a d x^b(d\ba{\q}\G^{c} d\q)+\frac{\ii}{14} y_{abc} x^a (\ba{\q}\G^{b} d\q)(d\ba{\q}\G^{c} d\q)
  +\frac{87\ii}{140}\c_{ab\a}d x^ad x^b d\q^\a \no \\ &+\frac{51}{70}\c_{ab\a}d x^a (\ba{\q}\G^{b} d\q) d\q^\a- \frac{9}{140}\c_{ab\a}d x^a (d\ba{\q}\G^{b} d\q) \q^\a+\frac{3}{70}\c_{ab\a} x^a (d\ba{\q}\G^{b} d\q) d\q^\a\no \\ &
  -\frac{\ii}{4}\c_{ab\a} (\ba{\q}\G^{a} d\q)(\ba{\q}\G^{b} d\q) d\q^\a +\frac{\ii}{28}\c_{ab\a} (\ba{\q}\G^{a} d\q)(d\ba{\q}\G^{b} d\q) \q^\a
  -\frac{9\ii}{35}w_{a\a\b} d x^a d\q^\a d\q^\b \no \\&+\frac{3}{140}w_{a\a\b} (d\ba{\q}\G^{a} d\q) \q^\a d\q^\b -\frac{3}{20}w_{a\a\b} (\ba{\q}\G^{a} d\q) d\q^\a d\q^\b-\frac{3\ii}{35}\f_{\a\b\g}d\q^\a d\q^\b d\q^\g\,.
\end{align}

Since $\mathcal{F}^{(4)}=d A^{(3)}-b^{(4)}$ is by construction manifestly invariant under supersymmetry transformations, it should be able to be expressed in terms of Maurer-Cartain forms associated with the enlarged supersymmetry algebra. Using the coset parametrisation (\ref{Omega}), the corresponding Maurer-Cartan forms are
\begin{multline}
 \W(x, \q, y, \c , w, \f)^{-1} d \W(x, \q, y, \c , w, \f)\\ =  \ii \left(E^a P_a+E^\a Q_\a+E_{abc}Z^{abc}+E_{ab\a}Z^{ab\a}+E_{a\a\b}Z^{a\a\b}+E_{\a\b\g}Z^{\a\b\g}  \right)
\end{multline}
with
\begin{align}
  E^a&=d x^a-\ii (\ba{\q}\G^a d \q)\\
  E^\a &= d \q^\a\\
  E_{abc}&=d y_{abc}+\frac{\ii}{2}(\ba{\q}\G_{abc} d \q)\\
  E_{ab\a}&=d \c_{ab\a}+\frac{\ii}{2}x^c(d\ba{\q}\G_{cab})_\a -\frac{\ii}{2}dx^c(\ba{\q}\G_{cab})_\a-\ii y_{abc}(d\ba{\q}\G^{c})_\a
  \no\\& \phantom{=}+\ii d y_{abc}(\ba{\q}\G^{c})_\a-\frac13 (\ba{\q}\G^{c} d \q)(\ba{\q}\G_{cab})_\a-\frac13 (\ba{\q}\G_{abc} d \q)(\ba{\q}\G^{c})_\a
\end{align}
\begin{align}
  E_{a\a\b}&=d w_{a\a\b}-\frac{\ii}{2}x^b dx^c(\G_{abc})_{\a\b}- 2 \ii d\c_{ab(\a}(\ba{\q}\G^{b})_{\b)} +2 \ii \c_{ab(\a}(d\ba{\q}\G^{b})_{\b)} \no \\
  & \phantom{=}-\frac{\ii}{4}d\c_{ab\g}\q^\g(\G^{b})_{\a\b} +\frac{\ii}{4}\c_{ab\g}d\q^\g(\G^{b})_{\a\b}
   +\frac{\ii}{2}x^b d y_{bca}(\G^{c})_{\a\b} \no\\& \phantom{=}-\frac{\ii}{2} y_{abc}dx^b(\G^{c})_{\a\b}-\frac13 x^b (\ba{\q}\G^{c} d \q)(\G_{abc})_{\a\b}-\frac12 y_{abc}(\ba{\q}\G^{b} d \q)(\G^{c})_{\a\b}\no\\& \phantom{=}-\frac{1}{12}x^b(\ba{\q}\G_{bca} d \q)(\G^{c})_{\a\b}
   -\frac23x^b(d\ba{\q}\G_{bca})_{(\a}(\ba{\q}\G^{c})_{\b)} +\frac23 dx^b(\ba{\q}\G_{bca})_{(\a}(\ba{\q}\G^{c})_{\b)} \no\\&\phantom{=}
    +\frac43 y_{abc}(d\ba{\q}\G^{b})_{(\a}(\ba{\q}\G^{c})_{\b)}-\frac43 dy_{abc}(\ba{\q}\G^{b})_{(\a}(\ba{\q}\G^{c})_{\b)}
    -\frac{\ii}{3}(\ba{\q}\G^{b}d \q)(\ba{\q}\G_{bca})_{(\a}(\ba{\q}\G^{c})_{\b)}\no\\&\phantom{=}-\frac{\ii}{3}(\ba{\q}\G_{abc}d \q)(\ba{\q}\G^{b})_{(\a}(\ba{\q}\G^{c})_{\b)}\,.
\end{align}
\begin{align}
    E_{\a\b\g}&= d\f_{\a\b\g}-\frac13 x^a y_{abc}(d\ba{\q}\G^{b})_{(\a}(\G^{c})_{\b\g)}
    -\frac13 dx^a y_{abc}(\ba{\q}\G^{b})_{(\a}(\G^{c})_{\b\g)}\no \\& \phantom{=}+\frac23 x^a dy_{abc}(\ba{\q}\G^{b})_{(\a}(\G^{c})_{\b\g)}
    +\frac{\ii}{3}y_{abc}(\ba{\q}\G^{a}d\q)(\ba{\q}\G^{b})_{(\a}(\G^{c})_{\b\g)}
    +\frac{5\ii}{6}y_{abc}(\ba{\q}\G^{a})_{(\a}(\ba{\q}\G^{b})_{\b}(d\ba{\q}\G^{c})_{\g)}\no \\& \phantom{=}-\frac{5\ii}{6}dy_{abc}(\ba{\q}\G^{a})_{(\a}(\ba{\q}\G^{b})_{\b}(\ba{\q}\G^{c})_{\g)}+\frac{\ii}{2}x^a d\c_{ab(\a}(\G^{b})_{\b\g)}
    -\frac{\ii}{2}dx^a \c_{ab(\a}(\G^{b})_{\b\g)}\no \\& \phantom{=}-\frac12 \c_{ab(\a}(\ba{\q}\G^{a}d \q)(\G^{b})_{\a\b)}
    +\frac53 \c_{ab(\a}(\ba{\q}\G^{a})_{\b}(d\ba{\q}\G^{b})_{\g)} -\frac53 d\c_{ab(\a}(\ba{\q}\G^{a})_{\b}(\ba{\q}\G^{b})_{\g)}\no \\& \phantom{=}
    +\frac16\q^\l \c_{ab\l}(d\ba{\q}\G^{a})_{(\a}(\G^{b})_{\b\g)}+\frac16 d\q^\l \c_{ab\l}(\ba{\q}\G^{a})_{(\a}(\G^{b})_{\b\g)}
    -\frac13 \q^\l d\c_{ab\l}(\ba{\q}\G^{a})_{(\a}(\G^{b})_{\b\g)}\no \\& \phantom{=}-\frac{\ii}{4}d \q^\l w_{a\l(\a}(\G^{a})_{\b\g)}
    +\frac{\ii}{4}\q^\l dw_{a\l(\a}(\G^{a})_{\b\g)}-\frac{5\ii}{4}w_{a(\a\b}(d\ba{\q}\G^{a})_{\a)}+\frac{5\ii}{4}dw_{a(\a\b}(\ba{\q}\G^{a})_{\a)}\no \\& \phantom{=}-\frac{1}{12}x^a d x^b(\ba{\q}\G_{abc})_{(\a}(\G^{c})_{\b\g)}+\frac{5}{12}x^a d x^b(\ba{\q}\G^{c})_{(\a}(\G_{abc})_{\b\g)}-\frac{5\ii}{24}x^a(\ba{\q}\G^{b}d\q)(\ba{\q}\G^{c})_{(\a}(\G_{abc})_{\b\g)}\no \\& \phantom{=}
    +\frac{\ii}{24}x^a(\ba{\q}\G^{b}d\q)(\ba{\q}\G_{abc})_{(\a}(\G^{c})_{\b\g)} +\frac{\ii}{12}x^a(\ba{\q}\G_{abc}d\q)(\ba{\q}\G^{b})_{(\a}(\G^{c})_{\b\g)}\no \\& \phantom{=}+\frac{5\ii}{12}dx^a(\ba{\q}\G^{b})_{(\a}(\ba{\q}\G^{c})_{\b}(\ba{\q}\G_{abc})_{\g)}
    -\frac{5\ii}{12}x^a(\ba{\q}\G^{b})_{(\a}(\ba{\q}\G^{c})_{\b}(d\ba{\q}\G_{abc})_{\g)}\no \\& \phantom{=}
    +\frac16(\ba{\q}\G_{abc}d\q)(\ba{\q}\G^{a})_{(\a}(\ba{\q}\G^{b})_{\b}(\ba{\q}\G^{c})_{\g)}
    +\frac16(\ba{\q}\G^{a}d\q)(\ba{\q}\G^{b})_{(\a}(\ba{\q}\G^{c})_{\b}(\ba{\q}\G_{abc})_{\g)}.
\end{align}

Using these results, we find the following expression for the supersymmetry invariant four-form $\mathcal{F}^{(4)}=d A^{(3)}-b^{(4)}$:
\begin{align}
  \mathcal{F}^{(4)}=\frac{\ii}{2}E_{abc}E^aE^bE^c+\frac{87 \ii}{140}E_{ab\a}E^a E^b E^\a-\frac{9 \ii}{35}E_{a\a\b}E^a E^\a E^\b-\frac{3\ii}{35}E_{\a\b\g}E^\a E^\b E^\g,
\end{align}
which is independent of the constants $m_1$, $m_2$ and $m_3$.

For completeness, we provide a summary of our results for the enlarged superspace in the case $p=2,$ based upon the results of Section \ref{p=2a} and \ref{p=2b}. This case was treated in Section 8 of \cite{Chryssomalakos:1999xd}, but our expression for $ A^{(2)}$ differs slightly from theirs\footnote{Note that we have used different conventions from those in \cite{Chryssomalakos:1999xd}.  Upon converting our expression for $A^{(2)}$ to those conventions, we find disagreement in the last three coefficients of their expression (8.11).   Our expression for $\mathcal{F}^{(3)}$, however, coincides precisely with theirs after conversion. The following fundamental replacements can be used to  translate our expressions into the notation of \cite{Chryssomalakos:1999xd}: $P_a\rightarrow -\frac12 X_\m$, $Q_\a \rightarrow D_\a$, $Z^{ab} \rightarrow Z^{\m\n}$, $Z^{a\a} \rightarrow -\frac12 Z^{\m\a}$, $Z^{\a\b} \rightarrow \frac14 Z^{\a\b}$, $x^a\rightarrow 2 \ii x^\m$, $\q^\a \rightarrow -\ii \q^\a$, $y_{ab} \rightarrow -\ii \vf_{\m\n}$, $\c_{a\a} \rightarrow 2 \ii \vf_{\m\a}$, $\f_{\a\b} \rightarrow - 4 \ii \vf_{\a\b}$.}, and is more general in the sense that it contains additional terms with arbitrary constants. The coset is based on a parametrization
\begin{align}
  \W(x, \q,\c,y,\f)=\exp\left\{\ii (x^a P_a+\q^\a Q_\a+\c_{a\a}Z^{a\a}+y_{ab}Z^{ab}+\f_{\a\b}Z^{\a\b} )\right\}\,,
\end{align}
and the corresponding Maurer-Cartan forms are
\begin{align}
  \W(x, \q,\c,y,\f)^{-1} d \W(x, \q,\c,y,\f)= \ii \left(E^a P_a+E^\a Q_\a+E_{a\a}Z^{a\a}+E_{ab}Z^{ab}+E_{\a\b}Z^{\a\b}  \right)\,.
\end{align}
We find that
\begin{align}
  A^{(2)}=\, &\, (\a-\tfrac{19}{30})x^a d x^b (\ba{\q}\G_{ab} d\q)+\ii(\l-\tfrac{11}{90})x^a(\ba{\q}\G^{b} d\q)(\ba{\q}\G_{ab} d\q)
  + \frac{2 \ii}{3} y_{ab}d x^a d x^b\no\\&+\frac{11}{15}y_{ab} dx^a(\ba{\q}\G^b d\q)
  +\frac{1}{15}y_{ab}x^a(d\ba{\q}\G^b d\q)-\frac{11\ii}{45}y_{ab}(\ba{\q}\G^a d\q)(\ba{\q}\G^b d\q)-\frac{3\ii}{5}\c_{a\a}dx^a d\q^\a
  \no\\&+\frac{1}{30}\c_{a\a}\q^{\a}(d\ba{\q}\G^a d\q)-\frac{1}{3}\c_{a\a}(\ba{\q}\G^a d\q)d\q^\a-\frac{2\ii}{15}\f_{\a\b}d\q^{\a}d\q^{\b}\,,\label{A}
\end{align}
and the manifestly supersymmetric three-form
\begin{align}
  \mathcal{F}^{(3)}=d A^{(2)}-b^{(3)}=\frac{2 \ii}{3}E_{ab}E^aE^b-\frac{3 \ii}{5}E_{a\a}E^aE^\a-\frac{2 \ii}{15}E_{\a\b}E^\a E^\b\,,
\end{align}
which is independent of the constants $\a$ and $\l$.

\section{Conclusions}
\setcounter{equation}{0}

In this paper, we have used integrability of the supersymmetry transformations assigned to $A^{(p)}$ in solving the  cohomology problem
\be
h^{(p+2)} = d b^{(p+1)}, \quad \d_{\e} b^{(p+1)} = d \d_{\e} A^{(p)}.
\ee
associated with the closed and supersymmetry invariant superspace $(p+2)$-forms $h^{(p+2)}$ to systematically derive the enlarged superalgebras that underly $p$-branes for the cases $p=2,$ $p=3,$ as well as the $D$-brane worldvolume one-form and the $M5$ brane two-form. We have also given a coset construction for the manifestly supersymmetric Wess-Zumino Lagrangian for a 3-brane.

In a future work we will generalise the procedure to compute the algebra for all valid $p$ and then use the result to compute manifestly supersymmetric expressions for $\mathcal{F}^{(p+1)}$.

\appendix
\section{Establishing equation \eqref{d12Aa}}\label{appendix}
\setcounter{equation}{0}
Here we provide further details on how the expression \eqref{d12Aa} along with the conditions \eqref{consts} emerge.

For convenience it is useful to introduce the following notation
\begin{align}
  P_1&:= dx^a dx^b(\ba{\e}_{[1}\G_{ab} \e_{2]})= dx^a dx^b(\ba{\e}_{1}\G_{ab} \e_{2})\\
  Q_1&:= x^a(\ba{\e}_{[1}\G^b \e_{2]})(d\ba{\q}\G_{ab}d\q) = x^a(\ba{\e}_{1}\G^b \e_{2})(d\ba{\q}\G_{ab}d\q)\\
  Q_2&:= x^a(d\ba{\q}\G^{b}d\q)(\ba{\e}_{[1}\G_{ab} \e_{2]})=x^a(d\ba{\q}\G^{b}d\q)(\ba{\e}_{1}\G_{ab} \e_{2})\\
  Q_3&:= x^a(\ba{\e}_{[1}\G^{b}d\q)(\ba{\e}_{2]}\G_{ab}d\q)\\
  R_1&:= dx^a(\ba{\e}_{[1}\G^b \e_{2]})(\ba{\q}\G_{ab} d \q )=dx^a(\ba{\e}_{1}\G^b \e_{2})(\ba{\q}\G_{ab} d \q )\\
  R_2&:= dx^a(\ba{\q}\G^{b}d\q)(\ba{\e}_{[1}\G_{ab} \e_{2]})=dx^a(\ba{\q}\G^{b}d\q)(\ba{\e}_{1}\G_{ab} \e_{2})\\
  R_3&:= dx^a(\ba{\e}_{[1}\G^{b}\q)(\ba{\e}_{2]}\G_{ab}d\ba{\q})\\
  R_4&:= dx^a(\ba{\e}_{[1}\G^{b}d\q)(\ba{\e}_{2]}\G_{ab}\q)\\
  S_1&:= (\ba{\e}_{[1}\G^a \e_{2]})(\ba{\q}\G^bd\q)(\ba{\q}\G_{ab}d\q) = (\ba{\e}_{1}\G^a \e_{2})(\ba{\q}\G^bd\q)(\ba{\q}\G^{ab}d\q)\\
  S_2&:= (\ba{\q}\G^ad\q)(\ba{\q}\G^{b}d\q)(\ba{\e}_{[1}\G_{ab} \e_{2]})  = (\ba{\q}\G^ad\q)(\ba{\q}\G^{b}d\q)(\ba{\e}_{1}\G_{ab} \e_{2})\\
  S_3&:= (\ba{\e}_{[1}\G^a\q)(\ba{\e}_{2]}\G^{b}\q)(d\ba{\q}\G_{ab} d \q )  =(\ba{\e}_1\G^a\q)(\ba{\e}_2\G^{b}\q)(d\ba{\q}\G_{ab} d \q ) \\
  S_4&:= (\ba{\e}_{[1}\G^a\q)(\ba{\e}_{2]}\G^{b}d\q)(\ba{\q}\G_{ab} d \q )\\
  S_5&:= (\ba{\e}_{[1}\G^a\q)(d\ba{\q}\G^{b}d\q)(\ba{\e}_{2]}\G_{ab} \q )\\
  S_6&:= (\ba{\e}_{[1}\G^ad\q)(\ba{\q}\G^{b}d\q)(\ba{\e}_{2]}\G_{ab} \q )\\
  S_7&:= (\ba{\e}_{[1}\G^a\q)(\ba{\q}\G^{b}d\q)(\ba{\e}_{2]}\G_{ab} d \q ),
\end{align}
which is an exhaustive list of all possible terms that could appear in $[\d_{\e_2},\d_{\e_1}] A^{(2)}$.  Some of these terms are related via the gamma matrix identity \eqref{gammaid3}, as follows:
\begin{align}
  Q_1+Q_2-4 Q_3&=0\,, \label{id1} \\
  R_1+R_2-2 R_3+2 R_4 &=0\,, \label{id2}\\
  S_3-2 S_4+S_5+2 S_7&=0\,, \label{id3} \\
  S_1-S_2-2 S_6-2 S_7 &=0\,. \label{id4}
\end{align}

By taking a second supersymmetry variation of \eqref{ssva} (with \eqref{c2} and \eqref{a1}) and forming the commutator, a direct calculation yields
\begin{align}
  [\d_{\e_2},\d_{\e_1}]& A^{(2)}= 2 \a  \b  P_1+2\ii(1-\a)(1-\g)Q_1
+2 \ii \k  Q_2 \no \\ &+ 2\ii\left( \a  \g -2\a - \g +\frac{5}{3}\right)R_1
+2 \ii\left(\k -\frac13\right) R_2
+2\left(\frac{1}{15}- \lambda \right)S_1
-\frac{2 }{15}  S_2\no \\ &
+ 2\ii\Big( \a  (\b-1)+  \g(\a -1)+ \k \Big)Q_3
 -2 \ii\left( \a  (\b -\g +1)+ \g -\frac53\right)R_3\no \\ &
- 2\ii \left(2\a  \b +\k -\frac13\right)R_4
+ 2(1- \a)(1-\g)S_3\no \\ &
-4\left( \a + \g - \a \g-\frac{13}{15}\right)S_4
+2 \k  S_5
+2\left(\k -\frac15\right)S_6
-2\left( \k -\frac{1}{15}\right) S_7 \,.
\end{align}
Here  we have arranged the terms so that only those containing a factor of $(\bar{\e}_{1}\G^{a}\e_2)$ or $(\ba{\e}_1\G_{ab}\e_2)$  -- the former relating to the standard supersymmetry algebra and the latter to  charges in the enlarged algebra -- appear on the first two lines.  We are now left with the problem of removing all other terms so that we are furnished with a realisation of the enlarged algebra.   We eliminate these unwanted terms by exploiting some of the freedom in the above expression due to the presence of the arbitrary constants $\a$, $\b$, $\g$ and $\l$ and the relationships \eqref{id1}-\eqref{id4}.  From this point it is not difficult to establish that the most general solution to this problem requires imposing \eqref{consts} which yields the result \eqref{d12Aa}.

\begin{footnotesize}

\end{footnotesize}

\end{document}